\documentclass{ptephy_v1}%%%%%% to generate preprint number with ptep logo

\preprintnumber{XXXX-XXXX} %%% %%% Insert preprint number here

\usepackage{graphicx}% Include figure files
\usepackage{dcolumn}% Align table columns on decimal point
\usepackage{bm}% bold math
\usepackage{mathrsfs}

\begin{document}

\title{
	Toward experimental observations of induced Compton scattering by high-power laser facilities
}

%%%% To generate auto affiliation numbers please use \author{}\affil{} command

\author[1]{Shuta J. Tanaka}
\affil{Department of Physics and Mathematics, Aoyama Gakuin University, 5-10-1 Fuchinobe, Sagamihara, Kanagawa 252-5258, Japan
\email{sjtanaka@phys.aoyama.ac.jp}
}

\author[1,2]{Ryo Yamazaki}
\affil{Institute of Laser Engineering, Osaka University, 2-6 Yamadaoka, Suita, Osaka 565-0871, Japan}

\author{Yasuhiro Kuramitsu}
% \author[3]{Insert fourth author name here} %%% Use optional bracket [3] to change the respective address
\affil{Graduate School of Engineering, Osaka University, 2-1 Yamadaoka, Suita, Osaka 565-0871, Japan}

\author[2]{Youichi Sakawa}

%%% To include the collaborator name... Please use the command "\collaborator"
%%% For example: \collaborator{ATLAS Collaboration}

\begin{abstract}%
    Induced Compton scattering (ICS) is a nonlinear interaction between intense electromagnetic radiation and a rarefied plasma.
    Although the magnetosphere of pulsars is a potential cite at which ICS occurs in nature, the ICS signatures have not been discovered so far.
    One of the reasons for non-detection of the ICS signatures is that we still do not attain the concrete understanding of such nonlinear plasma interactions because of their nonlinear nature and of the lack of experimental confirmations.
    Here, we propose a possible approach to understand ICS experimentally in laboratories, especially, with the use of the up-to-date short-pulse lasers.
    We find that the scattered light of ICS has characteristic signatures in the spectrum.
    The signatures will be observed in some current laser facilities.
    The characteristic spectrum is quantitatively predictable and we can diagnose the properties of the scattering plasma from the signatures.
\end{abstract}

\subjectindex{xxxx, xxx}

\maketitle

%%%%%%%%%%%%%%%%%%%%%%%%%%%%%%%%%%%%%%%%%
%%%%%%%%%%%%%%%%%%%%%%%%%%%%%%%%%%%%%%%%%
\section{Introduction}
%%%%%%%%%%%%%%%%%%%%%%%%%%%%%%%%%%%%%%%%%

Pulsars --- rotating neutron stars --- are laser transmitters in nature.
Their radio emission is coherent and their brightness temperature is so high, e.g., the Crab pulsar has record of $10^{37}$ K at 5.5 GHz \cite{Hankins+03}.
The mechanism of the coherent radio emission is a long-standing mystery after the unexpected discovery of a neutron star in radio wavelength \cite{Hewish+68} and is a challenge from the nature to our knowledge of quantum and relativistic plasma physics \cite{Melrose17}.
Here, we discuss a potential to understand one of plasma processes in such an extreme condition experimentally by using up-to-date short-pulse lasers.
The process which we study is known as induced Compton scattering (ICS).

\begin{figure}
\begin{center}
   	\includegraphics[scale=0.6]{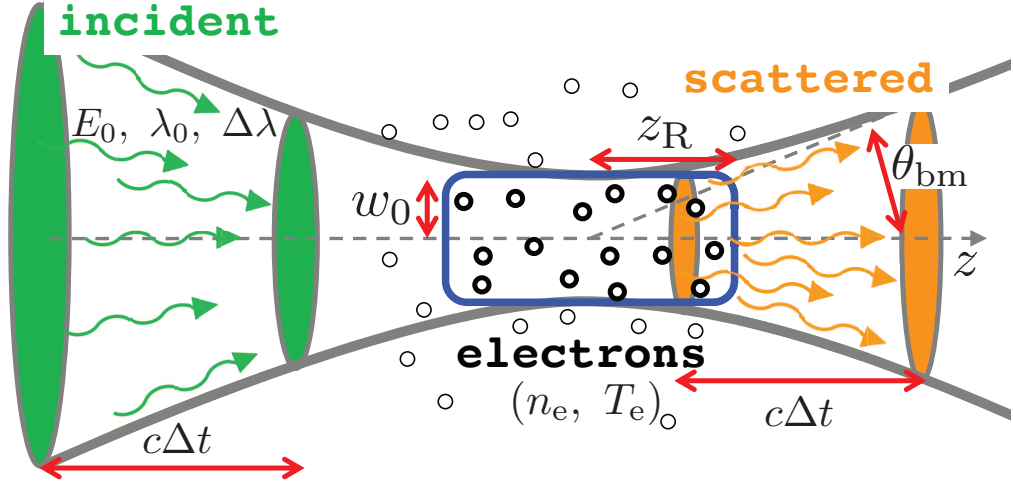}
\end{center}
\caption{
	Schematic picture of ICS in laboratory, i.e., an interaction between a Gaussian beam and electrons localized at the Rayleigh range (blue boxed region).
	The Rayleigh length $z_{\rm R}$ and the opening angle of the beam $\theta_{\rm bm}$ are written in the laser parameters of the incident light (see text).
}
\label{fig:laser}
\end{figure}

%ICS is a nonlinear interaction of strong electromagnetic waves and free electrons \cite{Dreicer64,Zeldovich75}, and is expected to occur around the pulsar magnetosphere \cite{Tanaka&Takahara13a}.
ICS is a nonlinear interaction of strong electromagnetic waves and free electrons \cite{Dreicer64,Zeldovich75}, and is expected to occur around the pulsar magnetosphere \cite{Tanaka&Takahara13a}.
As will be discussed in section \ref{sec:discussion}, ICS studied in the present paper is different from another nonlinear interaction of strong electromagnetic wave and free electrons, named nonlinear Thomson (Compton) scattering (NTS) \cite{Sarachik&Schappert70,Mackenroth&DiPiazza11,Seipt&Kampfer11,Bula+96,Chen+98,Yan+17}.
We have neither identified ICS signatures from pulsars nor in laboratories because we have not understood ICS quantitatively.
Recently, one of the authors predicted that the scattered photon spectrum has characteristic line-like structures at red-side of the incident spectrum \cite{Tanaka+15}.
The spectral signature contains the information of the plasma, so that the better understanding of ICS would allow us to study the properties of the pulsar magnetosphere with the unprecedented way.
Although their study assumed isotropic incident radiation field, we update their equation in order to apply laser radiation in laboratories, i.e., a directional narrow beam of an opening angle $\theta_{\rm bm}$ depicted in Fig. \ref{fig:laser}.

%%%%%%%%%%%%%%%%%%%%%%%%%%%%%%%%%%%%%%%%%
%%%%%%%%%%%%%%%%%%%%%%%%%%%%%%%%%%%%%%%%%
%%%%%%%%%%%%%%%%%%%%%%%%%%%%%%%%%%%%%%%%%

The formulation of ICS is based on the Boltzmann-Uehling-Uhlenbeck equation which is the kinetic equation taking into account quantum corrections \cite{Uehling&Uhlenbeck33}.
The laser radiation field is described by the occupation number, $n({\bm r},{\bm k})$, which is the photon number in a unit phase space volume $d^3 {\bm r} d^3 {\bm k} / (2 \pi)^3$ at $({\bm r}, {\bm k})$.
Laser radiation is usually described by a Gaussian beam, which is a solution of the paraxial Helmholtz equation deduced from the Maxwell's equations \cite{Kogelnik&Li66}.
We have to relate $n({\bm r}, {\bm k})$ with the five laser parameters: the total energy $E_0$, the central wavelength (frequency) $\lambda_0$ ($\nu_0$), the spectral (full-)bandwidth $\Delta \lambda$ ($\Delta \nu = c \Delta \lambda / \lambda^2_0$), the pulse width $\Delta t$ and the minimum beam waist $w_0$ (Fig. \ref{fig:laser}).
For simplicity, it is assumed that the incident radiation has the uniform directional distribution inside the solid angle $\Delta \Omega \approx \pi \theta^2_{\rm bm}$ ($\theta_{\rm bm} = \lambda_0 / \pi w_0 \ll 1$) with a Gaussian spectrum, and that it is spatially uniform at the Rayleigh range $-z_{\rm R} \le z \le z_{\rm R}$ (blue boxed region in Fig. \ref{fig:laser}), i.e.,
\begin{eqnarray}\label{eq:OccupationNumber}
	n(z,\nu)
    \approx
     n_0
%     e^{-2 \frac{(\nu-\nu_0)^2}{\Delta \nu^2}}~
     \exp\left(-2\frac{(\nu-\nu_0)^2}{\Delta \nu^2}\right),
    ~
    0 \le \theta \le \theta_{\rm bm}.
%    ~\nu_0 - \frac{\Delta \nu}{2} \le \nu \le \nu_0 + \frac{\Delta \nu}{2},
\end{eqnarray}
%
%% In addition, we set that the incident beam is homogeneous (independent from ${\bm r}$) in the Rayleigh range.
The normalization constant $n_0$ relates to the laser parameters with the use of the definition of the total photon energy $E_0 \equiv \int g h \nu n({\bm r},{\bm k}) d^3 {\bm r} d^3 {\bm k} / (2 \pi)^3$ as
\begin{eqnarray}\label{eq:Normalization}
	n_0
    \approx
    \frac{E_0}{h \nu_0 \Delta \nu \Delta t},
\end{eqnarray}
where we used $d^3 {\bm r} \approx \pi w^2_0 c \Delta t$, $d^3 {\bm k} / (2 \pi)^3 \approx \Delta \lambda \Delta \Omega / \lambda^4_0$, and the polarization degree of freedom $g = 1$.
Because the brightness temperature of the radiation is $k_{\rm B} T_{\rm b}(\nu) = h \nu n(\nu)$, Eq. (\ref{eq:Normalization}) corresponds to the Nyquist's relation.
Note that $T_{\rm b}$ is constant along the ray and the opening angle decreases with the distance from the Rayleigh range.
We set the opening angle is constant $\theta_{\rm bm}$ in the Rayleigh range and consider ICS at the Rayleigh range.
%% \cite{Rybicki&Lightman79}.
%% because the opening angle is important for ICS off the narrow beam (see below).

We consider ICS off laser radiation by free electrons of density $n_{\rm e}$ and temperature $T_{\rm e}$ localized at the Rayleigh range at which the occupation number $n(\nu)$ is spatially uniform.
In the low-temperature $\Theta \equiv k_{\rm B} T_{\rm e} / m_{\rm e} c^2 \ll 1$ and low-frequency $\lambda_{\rm e} / \lambda \equiv h \nu / m_{\rm e} c^2 \ll 1$ limits, the kinetic equation for photons becomes
\begin{eqnarray}\label{eq:ICS}
	\frac{\partial n}{\partial y}
	& = &
	\frac{\Theta \theta_{\rm bm}^6}{16} n
    \left\{
		(k_{\rm \Theta}^2 + 1) \frac{\partial}{\partial x} x^2 n
		+
		x \frac{\partial^3}{\partial x^3} x^3 n
	\right\}
	-
	\frac{n}{\Theta},
\end{eqnarray}
where $y \equiv n_{\rm e} \sigma_{\rm T} c t \Theta$ is the Compton $y-$parameter (characteristic time or length), $\sigma_{\rm T}$ is the Thomson cross section, $x \equiv h \nu / k_{\rm B} T_{\rm e}$ is a normalized photon frequency, and $k_{\Theta}^2 \equiv 3 / (2 \Theta \theta_{\rm bm}^2) - 1 \gg 1$ is a characteristic spectral width (see below) \cite{Tanaka+15}.
%% Derivation of Eq. (\ref{eq:ICS}) is straightforward but a long calculation along with our past study \cite{Tanaka+15}.
Eq. (\ref{eq:ICS}) is similar to the isotropic case \cite{Tanaka+15} but has an explicit strong dependence on $\theta_{\rm bm}$.
The first-derivative term in the right-hand side of Eq. (\ref{eq:ICS}) is the ``ICS term'' which is nonlinear in $n(x)$ and first order in terms of $\lambda_{\rm e} / \lambda (= \Theta x \propto xy$) and is corresponding to the quantum electron recoil effect boosted by the induced effect of bosons.
The third-derivative term, which we call the ``Doppler term'', is also nonlinear in $n(x)$ and is second-order in terms of $\Theta \lambda_{\rm e} / \lambda (= \Theta^2 x \propto x y \Theta$) but plays a crucial rule in order to avoid unphysical multi-valued solutions by the Doppler effect \cite{Tanaka+15}.
The remaining term represents Thomson scattering, i.e., zeroth-order spontaneous scattering ($\propto y / \Theta$), and then the higher-order spontaneous scattering effects can be neglected.

\begin{figure}
\begin{center}
 	\includegraphics[scale=0.8]{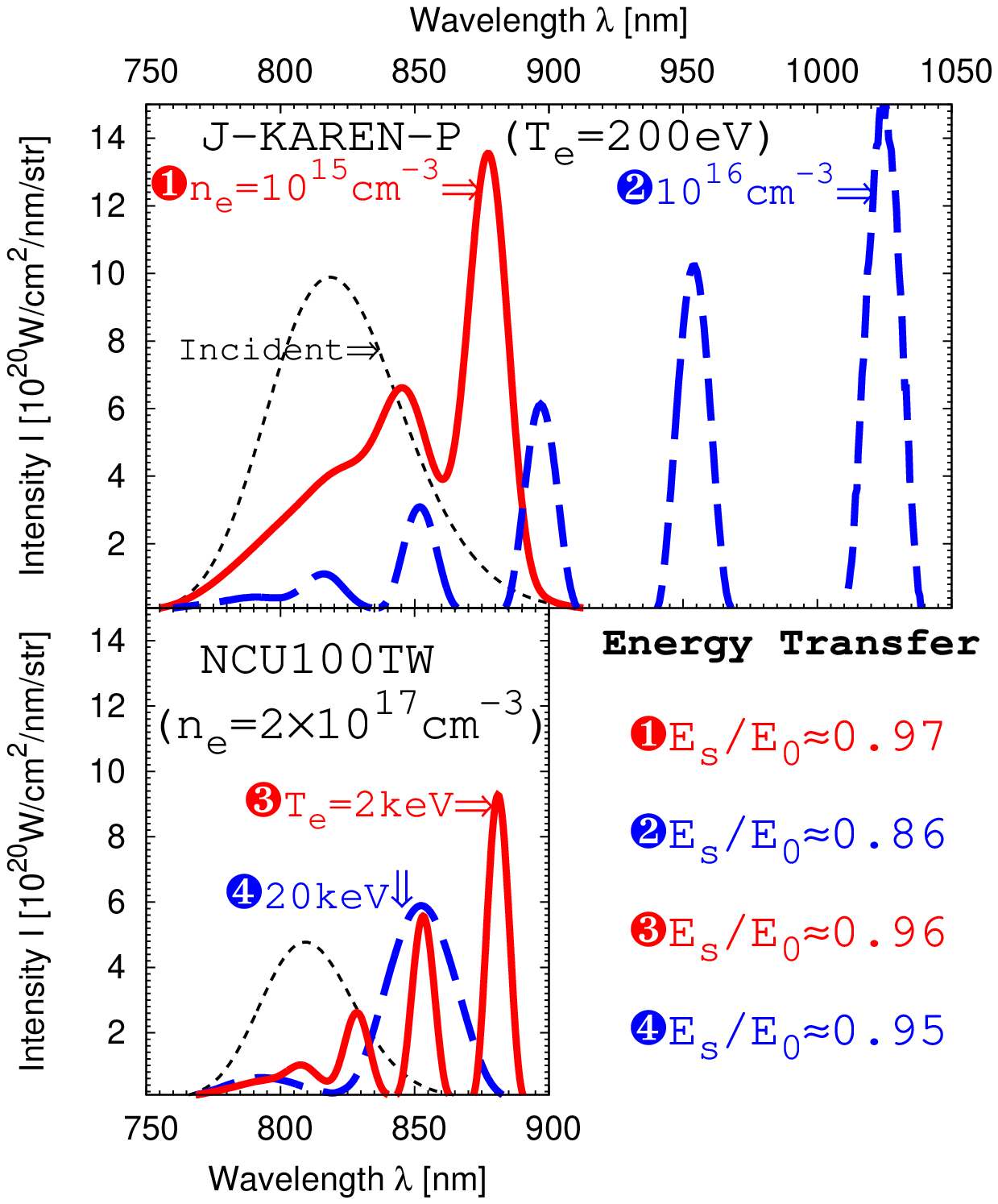}
\end{center}
\caption{
	Predicted spectra for laboratory experiments with J-KAREN-P (top) and by NCU100TW (bottom), where the dotted-black lines in both panels are the incident spectrum.
	(Top): $T_{\rm e} = 200$ eV ($\Delta \lambda k_{\Theta} / \pi \lambda_0 \approx 3.1$) is common while the thick-red and dashed-blue lines are $n_{\rm e} = 10^{15}~{\rm cm^{-3}}$ ($(\tau_{\rm ICS},\tau_{\rm D},\tau_{\rm Th}) \approx (0.09,10^{-5.5},10^{-12.6}$)) and $10^{16}~{\rm cm^{-3}}$ ($(\tau_{\rm ICS},\tau_{\rm D},\tau_{\rm Th}) \approx (0.9,10^{-4.5},10^{-11.6}$), respectively.
	(Bottom-left): $n_{\rm e} = 2 \times 10^{17}~{\rm cm^{-3}}$ ($(\tau_{\rm ICS},\tau_{\rm Th}) \approx (0.2,10^{-10.0})$) is common while the thick-red and dashed-blue lines are $T_{\rm e} = 2$ keV ($(\Delta \lambda k_{\Theta} / \pi \lambda_0, \tau_{\rm D}) \approx (1.4, 10^{-7.0})$) and 20 keV ($(\Delta \lambda k_{\Theta} / \pi \lambda_0,\tau_{\rm D}) \approx (4.5,10^{-6.0}$), respectively.
	(Bottom-right): The ratios of the total photon energy before and after ICS $E_{\rm s} / E_0$ are calculated for each case.
}
\label{fig:Predictions}
\end{figure}

%% The optical depth to each term of Eq. (\ref{eq:ICS}) is found from the dimension analysis $\tau \approx \partial n / n$.
%% Thomson scattering has the simplest form $\tau_{\rm Th} \equiv y / \Theta = n_{\rm e} \sigma_{\rm T} l$, where the scattering length $l = c t$ is set to twice the Rayleigh range $2 z_{\rm R} = 2 \pi w_0^2 / \lambda_0$.
The optical depth is given by $\tau \equiv \partial \ln n / \partial \ln y$, e.g., Thomson scattering has the well-known form $\tau_{\rm Th} \equiv y / \Theta = n_{\rm e} \sigma_{\rm T} l$, where the scattering length $l = c t$ is set to $2 z_{\rm R} = 2 \pi w_0^2 / \lambda_0$.
The optical depth to the ICS term is boosted by a factor of $\tau_{\rm ICS} / \tau_{\rm Th} \equiv (3/32) \theta_{\rm bm}^4 k_{\rm B} T_{\rm b} / m_{\rm e} c^2$ and has a strong dependence on $\theta_{\rm bm}$.
In terms of the laser parameters, we have
\begin{eqnarray}
	\frac{\tau_{\rm ICS}}{n_{\rm e}}
	=
	\frac{3 \sigma_{\rm T}}{16 \pi^3}
	\frac{E_0}{\Delta t \Delta \nu m_{\rm e} c^2}
	\frac{\lambda_0^3}{w_0^2}.
\end{eqnarray}
Finally, the optical depth to the Doppler term is $\tau_{\rm D} / \tau_{\rm Th} \equiv (\Theta/16) \theta_{\rm bm}^6 k_{\rm B} T_{\rm b} / m_{\rm e} c^2$.
The dispersive effect due to the third-derivative is the direct outcome of the Doppler term and then the ICS signatures will be observed for $\tau_{\rm ICS} \gtrsim 0.1 \gg \tau_{\rm D} > \tau_{\rm Th}$ (Fig. \ref{fig:Predictions}).
Typical values for some laser facilities are tabulated in Table \ref{tbl:LaserParameters}.

\begin{table}[b]%The best place to locate the table environment is directly after its first reference in text
\begin{minipage}{\textwidth}
	\caption{%
	Parameters of the several laser facilities: ``J-KAREN-P'' at National Institute for Quantum and Radiological Science and Technology in Japan \cite{Kiriyama+18}, ``NCU100TW'' at National Central University in Taiwan \cite{Hung+14}, and ``LFEX'' at osaka university in japan \cite{Kawanaka+08,Arikawa+16}.
	}
	\label{tbl:LaserParameters}
% \begin{ruledtabular}
\begin{center}
\begin{tabular}{c|lll}
	Parameters &
	J-KAREN-P  &
	NCU100TW   &
	LFEX \\
% \colrule
\hline
 	$E_0$ [J]             & 10   & 3.3 & 400  \\
 	$\lambda_0$ [nm]      & 820  & 810 & 1053 \\
 	$\Delta \lambda$ [nm] & 50   & 35  & 3.3  \\
 	$\Delta t$ [fs]       & 30   & 30  & 1500 \\
 	$w_0$ [$\mu$m]        & 0.67 & 4.3 & 50  \\
\hline
    $k_{\rm B} T_{\rm b} / m_{\rm e} c^2$        & 1.8 $\times$ 10$^{14}$  & 8.4 $\times$ 10$^{13}$  & 3.6 $\times$ 10$^{15}$  \\
    $\Delta \lambda / \lambda_0$                 & 6.1 $\times$ 10$^{-2}$  & 4.3 $\times$ 10$^{-2}$  & 3.1 $\times$ 10$^{-3}$  \\
	$\theta_{\rm bm}$                            & 3.9 $\times$ 10$^{-1}$  & 6.0 $\times$ 10$^{-2}$  & 1.3 $\times$ 10$^{-2}$  \\
	$z_{\rm R}$ [$\mu$m]                         & 1.7                     & 72                      & 1.9 $\times$ 10$^{ 3}$  \\
	$\tau_{\rm ICS}/n_{\rm e}$ [cm$^3$]          & 9.0 $\times$ 10$^{-17}$ & 9.7 $\times$ 10$^{-19}$ & 2.7 $\times$ 10$^{-18}$ \\
	$\tau_{\rm D} / (n_{\rm e} \Theta$) [cm$^3$] & 9.0 $\times$ 10$^{-18}$ & 2.3 $\times$ 10$^{-21}$ & 3.3 $\times$ 10$^{-22}$ \\
	$\tau_{\rm Th} /n_{\rm e} $[cm$^3$]          & 2.3 $\times$ 10$^{-28}$ & 9.5 $\times$ 10$^{-27}$ & 2.5 $\times$ 10$^{-25}$ \\
\end{tabular}
\end{center}
\end{minipage}
% \end{ruledtabular}
\end{table}

%

%%%%%%%%%%%%%%%%%%%%%%%%%%%%%%%%%%%%%%%%%
%%%%%%%%%%%%%%%%%%%%%%%%%%%%%%%%%%%%%%%%%
\section{Results}\label{sec:results}
%%%%%%%%%%%%%%%%%%%%%%%%%%%%%%%%%%%%%%%%%

Fig. \ref{fig:Predictions} shows numerical solutions to Eq. (\ref{eq:ICS}) applying the laser parameters of J-KAREN-P (top) and of NCU100TW (bottom-left).
For example, in the top panel of Fig. \ref{fig:Predictions}, we show the dependence on $\tau_{\rm ICS}$ so that $T_{\rm e} = 200$ eV is common while $n_{\rm e} = 10^{15}~{\rm cm^{-3}}$ ($\tau_{\rm ICS} \approx 0.09$) for thick-red and $n_{\rm e} = 10^{16}~{\rm cm^{-3}}$ ($\tau_{\rm ICS} \approx 0.9$) for dashed-blue lines.
The incident spectrum is a Gaussian (dotted-black line).
Note that the scattered spectrum is almost identical to the incident one for $\tau_{\rm ICS} \le 0.01$. 
As seen in the isotropic case \cite{Tanaka+15}, the scattered spectrum shows the line-like features in red-side of $\lambda_0$ and their width is predictable from the steady-state solution to Eq. (\ref{eq:ICS}),
\begin{eqnarray}\label{eq:SteadyStateSolution}
	x^2 \overline{n}(x)
	& = &
	A \cos (k_{\Theta} \ln x + \phi) + B + \frac{32}{3 \Theta \theta_{\rm bm}^4} x,
\end{eqnarray}
where an amplitude $A$, a DC component $B$, and a phase $\phi$ are constants of integration.
The first term in the right-hand side of Eq. (\ref{eq:SteadyStateSolution}) shows that characteristic spectral full-width is logarithmic, i.e., $\delta x / x  = \delta \lambda / \lambda \approx \pi / k_{\Theta}$, which is about $\theta_{\rm bm}$ times narrower than the isotropic case \cite{Tanaka+15}.
Taking $A = 0$, Eq. (\ref{eq:SteadyStateSolution}) becomes the steady-state solution without the Doppler (dispersive) term.

The spectra in Fig. \ref{fig:Predictions} show the five characteristics still similar to the isotropic case \cite{Tanaka+15}.
(1) The line-like features are formed intermittently and shift to longer wavelengths for larger $n_{\rm e}$ ($\tau_{\rm ICS}$), i.e., photons always lose their energy by ICS because of the electron recoil.
(2) The line-like features have a characteristic logarithmic width of $\pi / k_{\Theta}$.
(3) The number of the line-like feature increases with $n_{\rm e}$ and decreases with $T_{\rm e}$.
(4) The separation between the line-like features increases toward longer wavelengths.
(5) The intensity of the line-like features is higher at longer wavelengths.
Although the number of photons are conserved for ICS, the total energy of scattered photons $E_{\rm s}$ is smaller than that of incident ones $E_0$ (bottom-right of Fig. \ref{fig:Predictions}).

\begin{figure}[t]
\begin{center}
 	\includegraphics[scale=0.8]{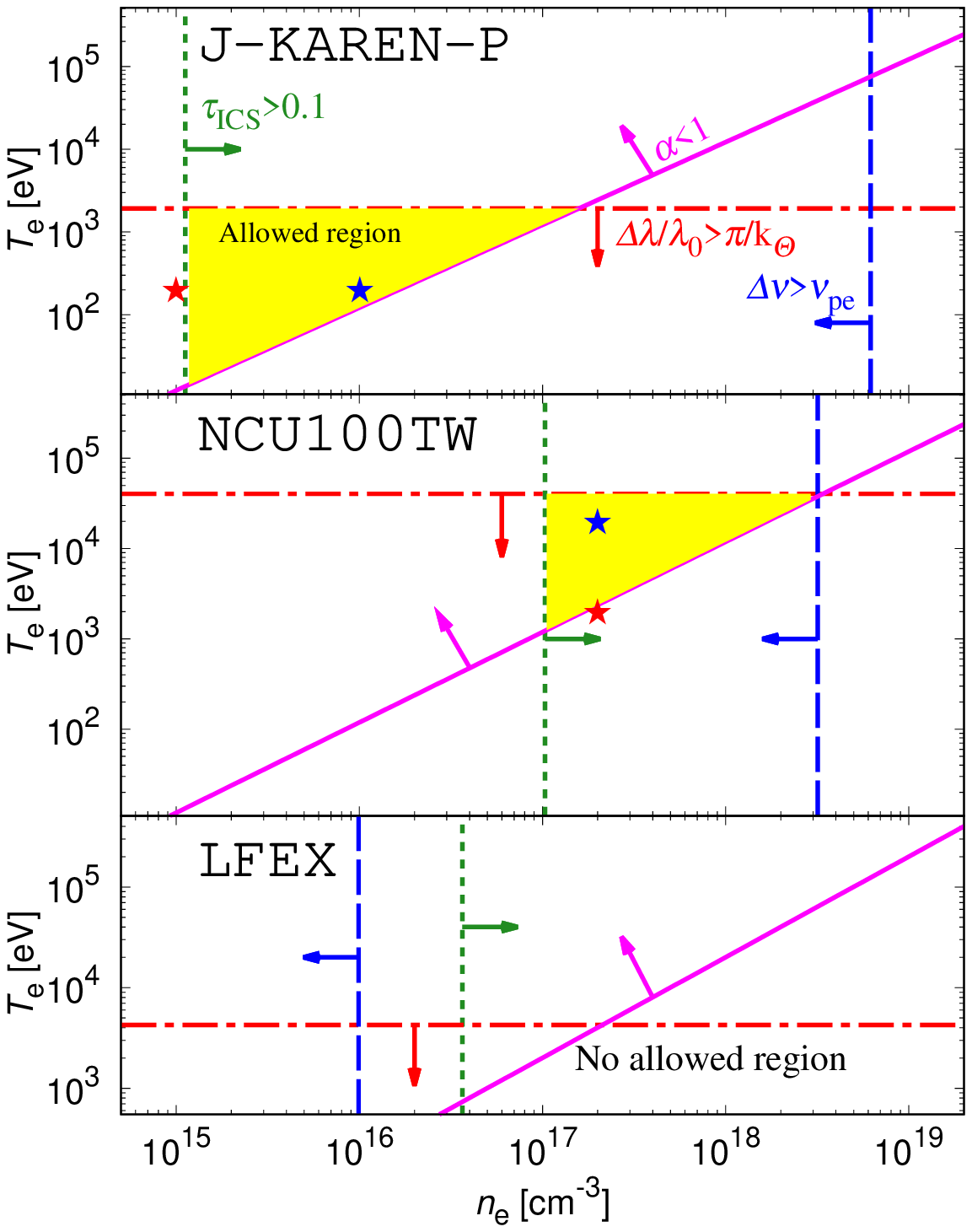}
\end{center}
\caption{
    Allowed regions (yellow marked) on the $n_{\rm e}-T_{\rm e}$ plane in order to observe the ICS signatures in each laser facility.
    We impose four necessary conditions (dotted-green, dot-dashed-red, dashed-blue and solid-magenta lines).
    The electron density should be large enough to deform the incident spectrum by ICS (Eq. (\ref{eq:Condition1})), while $n_{\rm e}$ should be small enough not to be contaminated by the plasma collective effects  (Eqs. (\ref{eq:Condition2}) and (\ref{eq:Condition4})).
    The electron temperature should be so small that the width of the line-like structure is smaller than the incident spectral width (Eq. (\ref{eq:Condition3})), while $T_{\rm e}$ should be large enough not to be contaminated by the plasma collective effects (Eq. (\ref{eq:Condition4})).
    Red and blue stars on the upper (J-KAREN-P) and middle (NCU100TW) panels are the parameters adopted in Fig. \ref{fig:Predictions}.
}
\label{fig:Constraints}
\end{figure}

Although $\tau_{\rm ICS}/n_{\rm e}$ (Table \ref{tbl:LaserParameters}) of LFEX is larger than that of NCU100TW, we currently do not expect to observe the ICS signatures with the use of LFEX.
This is because three additional conditions are imposed for the scattering electrons in order to observe the above five ICS signatures.
We put four necessary conditions below, and Fig. \ref{fig:Constraints} shows the allowed regions on the $n_{\rm e}-T_{\rm e}$ plane for given laser facilities.
The present ICS experiment is suitable for high-power short-pulse lasers (J-KAREN-P and NCU100TW) rather than high-energy (long-pulse) lasers (LFEX).

The first condition is $\tau_{\rm ICS} > 0.1$ (dotted-green lines in Fig. \ref{fig:Constraints}) and gives a lower limit on $n_{\rm e}$, such as
%% in order to satisfy (see Fig. \ref{fig:Predictions}) such as
\begin{eqnarray}\label{eq:Condition1}
    n_{\rm e}
    & > &
    1.1 \times 10^{15}~{\rm cm^{-3}}
    \left( \frac{E_0           }{10~{\rm J}} \right)^{-1}
    \left( \frac{\lambda_0     }{820~{\rm nm}} \right)^{-5}
    \left( \frac{\Delta t      }{30~{\rm fs}} \right)
    \left( \frac{\Delta \lambda}{50~{\rm nm}} \right)
    \left( \frac{w_0           }{0.67 \mu{\rm m}} \right)^{2}.
%%     \frac{1.6 \pi^3}{3 \sigma_{\rm T}}
%%     \frac{m_{\rm e} c^2 \Delta t \Delta \nu}{E_0}
%%     \frac{w^2_0}{\lambda^3_0}.
\end{eqnarray}
The second condition $\Delta \lambda / \lambda_0 > \pi / k_{\Theta}$ (dot-dashed-red) reduces an upper limit on $T_{\rm e}$ \cite{Tanaka+15} and is written as
\begin{eqnarray}\label{eq:Condition2}
     T_{\rm e}
     & < &
     31~{\rm keV}
    \left( \frac{\lambda_0     }{820~{\rm nm}} \right)^{-2}
    \left( \frac{\Delta \lambda}{50~{\rm nm}} \right)
    \left( \frac{w_0           }{0.67~\mu{\rm m}} \right).
%%      \sqrt{\frac{3}{2}} m_{\rm e} c w_0 \Delta \nu.
\end{eqnarray}
Note that the second condition is a practical requirement in order to observe the spectrum like Fig. \ref{fig:Predictions} \cite{Tanaka+15}.
Although the interaction which violates only Eq. (\ref{eq:Condition2}) is still ICS, its signatures will be different from those described above. 
The third (dashed-blue) and forth (solid-magenta) conditions are requirements in order to dismiss the plasma collective effects.
The higher-order Kompaneets equation (Eq. (\ref{eq:ICS})) is valid when the spectral width $\Delta \nu$ are greater than the Langmuir plasma frequency $\nu_{\rm pe} = \sqrt{n_{\rm e} e^2 / \pi m_{\rm e}}$ \cite{Galeev&Syunyaev73}, which reads
\begin{eqnarray}\label{eq:Condition3}
    n_{\rm e}
    & < &
    6.2 \times 10^{18}~{\rm cm^{-3}}
    \left( \frac{\lambda_0     }{820~{\rm nm}} \right)^{-4}
    \left( \frac{\Delta \lambda}{50~{\rm nm}} \right)^2.
%%     \frac{\pi m_{\rm e} \Delta \nu^2}{e^2}.
\end{eqnarray}
Finally, we require the scattering parameter $\alpha \equiv \lambda_0 / \lambda_{\rm De}$ is less than unity in order to avoid the effects of the plasma screening, where $\lambda_{\rm De} = \sqrt{k_{\rm B} T_{\rm e} / 4 \pi n_{\rm e} e^2}$ is the Debye length \cite{Evans&Katzenstein69}, that is,
\begin{eqnarray}\label{eq:Condition4}
     \frac{T_{\rm e}}{n_{\rm e}}
     & > &
     1.2 \times 10^{-17}~{\rm keV~cm^{3}}
    \left( \frac{\lambda_0     }{820~{\rm nm}} \right)^2.
%%      4 \pi e^2 \lambda^2_0 n_{\rm e}.
\end{eqnarray}
For high-energy (long-pulse) lasers such as LFEX, the condition of $\Delta \nu > \nu_{\rm e}$ (dashed-blue) contradicts the condition of $\tau_{\rm ICS} > 0.1$ (dotted-green).

%%%%%%%%%%%%%%%%%%%%%%%%%%%%%%%%%%%%%%%%%
%%%%%%%%%%%%%%%%%%%%%%%%%%%%%%%%%%%%%%%%%
\section{Discussion}\label{sec:discussion}
%%%%%%%%%%%%%%%%%%%%%%%%%%%%%%%%%%%%%%%%%

There have been some attempts to observe ICS in laboratories in the last century \cite{Krasyuk+70, Decroisette+72, Drake+90, Leemans+91, Everett+95}.
Based on the qualitative arguments done at those times \cite{Peyraud68a, Drake+74}, they reported some signatures of ICS.
%% Their ICS signatures are different from our prediction (five characteristics) and they made ICS experiments outside the allowed region on the $n_{\rm e}-T_{\rm e}$ plane discussed in Fig. \ref{fig:Constraints}.
%% We consider that their experiments would be contaminated by other phenomena.
%% The four conditions of the $n_{\rm e}-T_{\rm e}$ plane is imposed in order to observe pure ICS signatures.
%% Our prediction is rather quantitative and peculiar so that the observed signature can be distinguished from the other nonlinear laser-plasma interactions
However, none of those facilities has the allowed parameter region on the $n_{\rm e}-T_{\rm e}$ plane as in the case of LFEX (bottom panel of Fig. \ref{fig:Constraints}) and then the phenomena which they found would be different from `ICS' as we discussed in the present paper.
In addition, our prediction is rather quantitative and peculiar so that the observed signature can be distinguished from the other nonlinear laser-plasma interactions.

ICS studied in the present paper is considered as the different physical process from NTS by the following three reasons.
(1) ICS itself is the process preserving photon number \cite{Tanaka+15}, while NTS is not because, in the quantum view of NTS, multiple photons are absorbed by an electron and then the electron emits a single photon \cite{Seipt&Kampfer11,Mackenroth&DiPiazza11}.
(2) NTS is predominantly up-scattering process in contrast to ICS \cite{Sarachik&Schappert70,Chen+98} and (3) the scattering optical depths to ICS and NTS are not the same as follows.
The (effective) cross section of NTS, i.e., the optical depth to NTS, is boosted by the square of the laser strength parameter compared with (linear) Thomson scattering, i.e., $\tau_{\rm NTS} / \tau_{\rm Th} \approx a_0^2$, where $a_0^2 \equiv  (e \mathscr{E} \lambda_0 / 2 \pi m_{\rm e} c^2)^2 \propto \lambda^2_0 E_0 / \Delta t  w_0^2$ and $\mathscr{E}$ is the amplitude of the electric field.
Adopting the expression of the NTS cross section in classical limit from eq. (4.22) of \cite{Sarachik&Schappert70}, we have
\begin{eqnarray}\label{eq:NTS}
     \frac{\tau_{\rm ICS}}{\tau_{\rm NTS}}
     & = &
     \frac{3}{7 \sqrt{2} \pi^2} \frac{\lambda_0^4}{r_{\rm e} w_0^2 \Delta \lambda}
     \approx
     2.2 \times 10^8
    \left( \frac{\lambda_0     }{   820~{\rm nm}} \right)^{ 4}
    \left( \frac{\Delta \lambda}{    50~{\rm nm}} \right)^{-1}
    \left( \frac{w_0           }{0.67~\mu{\rm m}} \right)^{-2},
\end{eqnarray}
where the scattering length is also $l = 2 z_{\rm R}$ for NTS.
%because $a_0$ becomes the peak value at the focus.
Both ICS and NTS are clearly nonlinear processes because both $\tau_{\rm ICS}$ and $\tau_{\rm NTS}$ are proportional to the laser radiation energy flux $E_0 / \pi w_0^2 \Delta t$, however, they have different dependence on $\lambda_0,~\Delta \lambda$, and $w_0$ as seen in Eq. (\ref{eq:NTS}).
They are distinct physical processes and are experimentally distinguishable.

Although ICS seems to always dominate over NTS from Eq. (\ref{eq:NTS}), previous experiments on NTS were not suitable for observing the signatures of ICS \cite{Bula+96,Chen+98,Yan+17}.
For the experiments using the relativistic electron beam \cite{Bula+96,Yan+17}, Eq. (\ref{eq:NTS}) is modified by the relativistic Doppler effect.
The strength parameter $a_0$, i.e., $\tau_{\rm NTS} / \tau_{\rm Th}$, is an Lorentz invariant quantity but $\tau_{\rm ICS} / \tau_{\rm Th}$ can be very small depending on the direction of the electron beam with respect to the laser radiation \cite{Tanaka&Takahara13a}.
The experimental setup of \cite{Chen+98} is rather similar to the present study but has no allowed region on the $n_{\rm e} - T_{\rm e}$ plane (Fig. \ref{fig:Constraints}) for their laser parameters, where $\Delta t \Delta \nu = 1$ is assumed for deducing the spectral width from their paper.
%First, Eq. (\ref{eq:NTS}) is modified by the relativistic Doppler effect for the experiments using the relativistic electron beam \cite{Bula+96,Yan+17} because $a_0$, i.e., $\tau_{\rm NTS} / \tau_{\rm Th}$, is an Lorentz invariant quantity but $\tau_{\rm ICS} / \tau_{\rm Th}$ can be very small depending on the direction of the electron beam with respect to the laser radiation.
In addition, NTS has an experimental advantage against ICS that is the spectral signatures of NTS, i.e., higher harmonics of the incident radiation \cite{Sarachik&Schappert70,Seipt&Kampfer11,Mackenroth&DiPiazza11}, can be observed even for $\tau_{\rm NTS} \ll 1$.
For $\tau_{\rm NTS} \ll 1$, the scattered radiation is dimmer about $\tau_{\rm NTS}$ times than the (extremely bright) incident one but still shows the higher harmonic spectrum as long as $a_0 \gtrsim 1$.
For example, the observation of the second and third harmonics was made by \cite{Chen+98} with the optical depth to NTS of $\tau_{\rm NTS,Chen98} \sim 2 z_{\rm R} \sigma_{\rm T} a_0^2 n_{\rm e} \approx 10^{-6}$ for their fiducial electron density of $n_{\rm e} = 6.2 \times 10^{19}~{\rm cm^{-3}}$.
% Note that $\tau_{\rm ICS}/\tau_{\rm NTS} \gtrsim 0.1$ as seen in section \ref{sec:results}, but those of NTS are detectable for $\tau_{\rm NTS} \ll 1$ \cite{Chen+98}.
On the other hand, the spectral signatures of ICS, i.e, the line-like features of the similar intensity as the incident radiation (Fig. \ref{fig:Predictions}), do not appear for $\tau_{\rm ICS} < 0.1$ at all (Eq. (\ref{eq:Condition1})).
Although ICS far dominates over NTS in the present parameters (Eq. (\ref{eq:NTS})), because $a_0$ of J-KAREN-P is as high as a hundred, we have chance to observe the signatures of NTS simultaneously with those of ICS (Fig. \ref{fig:Predictions}).
The NTS signatures will be the $l$-th ($l \ge 2$) harmonics of the incident radiation ($\lambda_0 / l =$ 410 nm, 273 nm, $\cdots$) and their intensity will be less than $\sim 10^{13}~{\rm W~cm^{-2}~nm^{-1}~{str}^{-1}}$, i.e., about $\tau_{\rm NTS}$ times dimmer than the incident radiation.
% Even for $\tau_{\rm NTS} \ll 1$, the spectral signatures of NTS could be observed in addition to those of ICS, and those of NTS will have very small peaks at higher harmonics of the incident radiation \cite{Chen+98}, e.g., 420 nm, 205 nm, and so on, for J-KAREN-P.

Spectra of the scattered light of ICS inform us of the scattering electrons through $n_{\rm e}$ and $T_{\rm e}$.
An intriguing observational example in astrophysics is the giant radio pulses from the Crab pulsar.
Although they basically have broad spectra \cite{Mikami+16}, some show the spectral structure called `zebra band' which is the line-like structures of $\Delta \nu / \nu \approx 0.06$ discovered from the Crab pulsar's radio emission \cite{Hankins&Eilek07} and which can be a clue to understanding the physical condition of the pulsar magnetosphere.
We need to extend the model including the relativistic and the strong magnetic field effects for the interpretation of the zebra band.
Nevertheless, experimental confirmation of ICS in the present formulation is a step to understand the pulsar physics.
%% Although ICS cannot be the pulsar emission mechanism itself because it is a scattering process, the detection of the IcS signatures from the pulsars give implications to the physics of radio emission from the pulsars.
%% From its pulse width of $W \sim 6 {\rm \mu m}$, the opening of the radio pulse emission is $2 \pi W / P \sim 10^{-3}$ at the light cylinder, where $P = 33$msec is the period of the Crab pulsar.
%% \cite{Tanaka&Takahara13a}.

Alternative experiments can be done by the use of multiple beams. 
Strong dependence of ICS on $\theta_{\rm bm}$ results from Eq. (\ref{eq:OccupationNumber}) in which we consider ICS at the Rayleigh range of one beam (Fig. \ref{fig:laser}).
An extreme case would be the counter propagating beams which effectively increases the opening angle and ICS will be more effective than the present case.
We leave the modeling of ICS in such a situation as a future study because even the isotropization of a directional electro-magnetic beam by ICS is under discussion \cite{Goldin+75, Zeldovich&Syunyaev75, Lyubarsky&Petrova96}.
ICS with LFEX will come under consideration for multiple beam experiments.

Finally, in the present treatment, we neglect back-reactions to plasma.
The back-reactions do not only mean the plasma collective effects to scattered photons but also heating-up and/or acceleration of electrons by ICS \cite{Galeev&Syunyaev73}.
In most of astrophysical situations, the radiation pressure force caused by electron scattering is evaluated by $\tau_{\rm Th}$ as the Eddington luminosity \cite{Eddington16}.
However, the induced radiation pressure would be much stronger than the spontaneous one in this case \cite{Levich72}.
Interestingly, $\tau_{\rm ICS}$ increases with the laser power without increasing $n_{\rm e}$ so that individual electrons can easily attain to relativistic energy.
For example of J-KAREN-P, the total number of electrons in the Rayleigh range volume is only $2 \pi w_0^2 z_{\rm R} n_{\rm e} \approx 5 \times 10^3$ for $n_{\rm e} = 10^{15}~{\rm cm^{-3}}$.
The total energy transferred to electrons is $E_0 - E_{\rm s} \approx 0.3$ J and then each electron has $0.4$ EeV on average.
This mechanism can be possibly used for electron acceleration for the study of particle physics.

%%%%%%%%%%%%%%%%%%%%%%%%%%%%%%%%%%%%%%%%%%%%%
%%%%%%%%%%%%%%%%%%%%%%%%%%%%%%%%%%%%%%%%%%%%%
\section*{Acknowledgment}
%%%%%%%%%%%%%%%%%%%%%%%%%%%%%%%%%%%%%%%%%%%%%
%%%%%%%%%%%%%%%%%%%%%%%%%%%%%%%%%%%%%%%%%%%%%

S.J.T. would like to thank Y. Ohira, H. Takabe, Y. Fukuda, F. Takahara and also the anonymous referee for useful discussions and helpful comments.
This work is supported by JSPS Grants-in-Aid for Scientific Research Nos. 17H18270 (ST), 15H02154, 17H06202 (YS) and 18H01232 (RY), by Aoyama Gakuin University-Supported Program ``Early Eagle Program'' (ST), and also by the joint research project of the Institute of Laser Engineering, Osaka University No. 2019B2-TANAKA (ST).
R.Y. and S.J.T. deeply appreciate Aoyama Gakuin University Research Institute for helping our research by the fund.

\bibliographystyle{ptephy}
\bibliography{2002-008-3j-shutajtanaka}% Produces the bibliography via BibTeX.

\end{document}